\vskip5pt

\centerline  {\bf Quantum number conservation: a tool in the design and analysis of high energy experiments}
\vskip5pt
\centerline  {D.J.Newman}
\vskip5pt
\centerline  {email: \it dougnewman276@gmail.com}

\vskip5pt
\beginsection Abstract

The Clifford Unification algebra $Cl_{7,7}$ is shown to 
provide a unified description of all the elementary fermions 
and hadrons in terms of seven binary quantum numbers. Four of these
are defined in existing theories, namely spin-direction, 
fermion/anti-fermion pairing and the two commuting generators in 
the SU(3) description of quark colour. A $Cl_{3,3}$ sub-algebra
integrates the $Cl_{1,3}$ space-time and Dirac algebras  
providing a new definition of time direction and 
identifying parity as a new quantum number. A further two
quantum numbers are related to the commuting generators of an 
SU(3) description of fermion generations.
All seven are conserved in the decays and interactions 
of charged particles, suggesting that quantum number 
conservation provides a useful tool in the design 
and interpretation of high energy experiments. 
This is especially relevant when neutral particles 
and parity are involved.

\vfill\eject

\beginsection \S1. Introduction

The success of the Standard Model (SM) can be attributed partly to the fact
that results of zero neutrino mass and small finite neutrino mass calculations 
should give similar results.  Nevertheless,
the algebraic structures of small mass and zero mass theories have
fundamental differences that should be taken in 
constructing the unified theories. These differences will not be
discussed here, where the finite mass of neutrinos
is incorporated in the formulation.
This work is based on extending the accepted physical 
applications of Clifford algebras to space-time geometry [1,2] and 
Dirac's theory of the electron, which employ distinct interpretations 
of $Cl_{1,3}$. The idea that such a connection should exist was first
explored by Eddington [3], but his work was based on too little
experimental data to provide constraints on physical
interpretations of the algebra.

Recent attempts to develop algebraic theories of the elementary particles
have mostly accepted the SM as a starting point. Nevertheless, they
do provide important insights into
possible extensions of $Cl_{1,3}$ that describe the
properties of elementary fermions and bosons.
Trayling and Baylis [4,5] showed that the $Cl_{0,7}$ 
algebra reproduces many features
of the SM, but failed to find the 
extension necessary to develop it. Furey [6] showed that
some features of unified theories can be reproduced
by the structure of complex Clifford algebras.
Dartora and Cabrera [7] showed that the main 
features of electro-weak theory can be explained in terms of the $Cl_{3,3}$
algebra if chirality is omitted. Stoica [8] 
has shown that the results in [4,5,6] can also be expressed
using the complex Clifford algebra $Cl^*_6$, and has investigated 
how this algebra might incorporate chiral symmetry breaking.  
Yamatsu [9] has described a grand 
unified theory based on the Lie group USp(32), which is 
related to SO(32) string theory. Given that the Lie algebra
of SO(32) and $Cl_{5,5}$ are both described by
$32\times 32$ matrices, there are possible 
links between Yamatsu's work and the algebraic approach. 
Gording and Schmidt-May [10] developed a way to
integrate the Standard Model into the algebra 
of complex 8$\times$8 matrices, which is isomorphic
with the complex Clifford algebra $Cl^*_6$, but their
algebraic identifications of fermions bear no resemblance
to those developed in this paper. None of the above papers
identifies faults in the algebraic structure
of the Standard Model, or suggests that it should be
replaced. Most recently, Pav\v si\v c [11] has 
given string theoretic arguments that                             
$Cl_{8,8}$ is capable of providing a description of 
the elementary fermions and their interactions, 
but does not provide physical interpretations of 
the elements of this algebra.  Some recent work [12,13] 
has employed more complex algebraic structures that `go
beyond' the SM.  But, as these papers to build on the SM
they fail to provide solutions to the fundamental 
problems, viz. 
\item {(1)} How to reconcile the established applications
of $Cl_{1,3}$ in the Dirac equation and in space-time geometry.
\item {(2)} How to determine an algebraic characterisation
of time intervals.

This work demonstrates that the quantum number description of all 
the elementary fermions provided by Clifford Unification (CU)
algebra $Cl_{7,7}$ [14] has an application in the design and interpretation
of high energy experiments. While the CU formalism necessarily incorporates 
some features of the SM, it is based on distinct algebraic and 
conceptual structures. Detailed comparisons with the SM, detailed in [14,15], 
are omitted. Apart from the description of spinors,
explicit matrix representations are not required in the 
analyses but, for clarity, background details are given in Appendix B.
Content:
\item {\S2} Identifies $Cl_{3,3}$ as the algebra that integrates 
space-time geometry with Dirac's theory of the electron and gives an 
algebraic expression for time intervals. Three commuting elements  
define binary quantum numbers that describe fermion properties in space-time .
\item {\S3}Reformulates Dirac's theory so as to make its Lorentz invariance explicit.
\item {\S4}Reformulates weak interaction theory in terms of elements of $Cl_{3,3}$.
\item {\S5}Shows that two binary quantum numbers, defined by the additional commuting elements 
in $Cl_{5,5} \supset Cl_{2,2} $ algebra, distinguish leptons and quarks.
\item {\S6}Shows that commuting elements of $Cl_{7,7}$ define two further quantum numbers 
that distinguish the three known generations of fermions and describes their interactions.  
\item {\S7}Tabulates quantum number descriptions of all the elementary fermions.
\item {\S8}Provides examples of quantum number conservation in the decays and 
interactions of charged leptons and hadrons.
\item {\S9}Discusses the results obtained in this work and their potential applications. \vskip1pt
\item {Appendix A} Gives properties of Clifford algebras making them relevant to the description of fermions.
\item {Appendix B} Provides the matrix representation of $Cl_{3,3}$ relevant to the relationship between
space-time and Dirac algebras.

\beginsection \S2. Identification of $Cl_{3,3}$ as the fundamental algebra

$Cl_{1,3}$ provides an algebraic description of Minkowski
space-time. Its three space-like generators 
$\gamma^1,\gamma^2,\gamma^3$ describe
the properties of mutually orthogonal unit spatial displacements
in an observer's coordinate frame. 
Its time-like generator ($\gamma^a \equiv \gamma^0$)  
describes time intervals expressed in 
the same units as the spatial displacements.
Lorentz transformations between observer's frames 
can be described algebraically.
The physical applications of $Cl_{1,3}$ have 
been developed in considerable detail (e.g. [1,2]) 
without the need to employ matrix representations.
In contrast, the Dirac algebra is a specific 4$\times$4 matrix 
representation of $Cl_{1,3}$ that describes fermion 
properties and is not subject to Lorentz transformations.

The physical interpretation of Dirac's algebra provides the
starting point of this analysis.  Although it is seldom made
explicit in the literature, the $\gamma^\mu$ generators of
this algebra describe unit space-time displacements
in fermion rest frames. In the Dirac-Pauli matrix
representation, e.g. [18] equation (4.35), 
$\gamma^0$ and $\gamma^{12} = \gamma^1\gamma^2$ 
are diagonal. $\gamma^0$ is interpreted as the unit time 
interval, with eigenvalues that distinguish fermions and 
anti-fermions and specify the direction of time. The eigenvalues 
of $\gamma^{12}$ distinguish the two directions of
a spin orientated along the $\gamma^3$ axis (e.g. [18], ps. 86-89). 
In the case of electrons there are the four distinct solutions  
e$^-\!\!\uparrow$, e$^-\!\!\downarrow$, e$^+\!\!\uparrow$, 
e$^+\!\!\downarrow$, each of which corresponds to one of the 
four combinations of eigenvalues $\mu_0 = \pm 1,\>\mu_{12} = \pm i$ 
of the commuting elements $\gamma^0$ and ${\gamma}^{12}$.

The analysis in this and the following section identifies 
$Cl_{3,3}$ as the`fundamental algebra' that integrates 
the interpretations of the space-time and Dirac $Cl_{1,3}$ algebras. 
Following the nomenclature given in Appendix A, its generators 
are labelled $1,2,3,a,b,c$.
In order to distinguish elements of $Cl_{3,3}$ from 
those of $Cl_{1,3}$ they are written as $\hat\gamma$.
The three space-like elements, denoted $\hat\gamma^1,\>\hat\gamma^2,\>\hat\gamma^3$,
correspond to $\gamma^1,\>\gamma^2,\>\gamma^3$ in the space-time algebra, and
describe the unit spatial coordinate displacements. Unit spatial volumes
${\bf v}= \hat\gamma^1\hat\gamma^2\hat\gamma^3$ are invariant under 
rotations of the spatial coordinates, but not Lorentz transformations.

The star notation $\hat\gamma^{*\mu},\>\{\mu = 0,1,2,3\}$ 
distinguishes elements of the algebra in the 
fermion coordinate frame and correspond 
to the elements of the Dirac algebra. Stars are omitted in
$\hat\gamma^a\hat\gamma^b\hat\gamma^c$ as these time-like generators
are not subject to Lorentz transformations.  
The definition of time direction must
relate to the distinction between fermions and anti-fermions,
which is achieved if time intervals are defined as the 6-volume element of 
of $Cl_{3,3}$ defined, in any coordinate frame, as the product
$$
\hat\gamma_0= \hat\gamma^0 =\hat\gamma^1\hat\gamma^2\hat\gamma^3\hat\gamma^a\hat\gamma^b\hat\gamma^c
= {\bf v}\hat\gamma^\pi =-\hat\gamma^\pi {\bf v}.                          \eqno (2.1)
$$
Here $\hat\gamma^\pi=\hat\gamma^a\hat\gamma^b\hat\gamma^c$ and the volume element ${\bf v}= \hat\gamma^1\hat\gamma^2\hat\gamma^3 $.
This definition ensures that $\hat\gamma_0$ is time-like and 
anti-commutes with all six generators.  Unit space-time volumes, 
defined as
$$
\hat\gamma^\pi= \hat\gamma^0\hat\gamma^1\hat\gamma^2\hat\gamma^3 = \hat\gamma^0 {\bf v} = \hat\gamma^a\hat\gamma^b\hat\gamma^c.                                  \eqno (2.2)
$$
are Lorentz invariant. 

Any pair product of generators commutes with $\hat\gamma_0$ and all the
generators not included in the pair. Physical interpretations as fermion properties
are based on defining generators in the fermion coordinate frame and 
selecting three specific pair products, namely $\hat\gamma^{*13},
\>\hat\gamma^{*2a},\>\hat\gamma^{*bc}=\hat\gamma^{bc}$. The products 
$$
\hat\gamma^{13}\hat\gamma^{2a},\>\> \hat\gamma^{2a}\hat\gamma^{bc}
,\>\> \hat\gamma^{bc}\hat\gamma^{13},\>\>
\hat\gamma^{13}\hat\gamma^{2a}\hat\gamma^{bc}=\hat\gamma^{0}                  \eqno (2.3)
$$
commute with all the generators.  Physical interpretations are
based on quantum numbers A, B and C defined by the eigenvalues of 
$$
\hat\gamma^{\rm A}=i\hat\gamma^{*13},\>\>\hat\gamma^{\rm B}=\hat\gamma^{*0},
\>{\rm and}\>\>\hat\gamma^{\rm C}=i\hat\gamma^{bc}=i\hat\gamma^{*\pi a}.       \eqno (2.4)
$$ 
These definitions ensure that their eigenvalues, denoted A, B and C, are
all $\pm1$. The matrix representation of $Cl_{3,3}$ given in
Appendix B expresses the commuting elements 
$\hat\gamma^{*13},\>\hat\gamma^{*2a},\>\hat\gamma^{\pi a}$ 
as 8$\times$8 diagonal matrices. ($\hat\gamma^\pi$ is Lorentz invariant, so that 
stars are omitted here and in the table in Appendix B.) 

The pair products 
$\hat\gamma^{bc}=\hat\gamma^{\pi a},\>\hat\gamma^{ab}=\hat\gamma^{\pi c},\>\hat\gamma^{ca}=\hat\gamma^{\pi b}$
all commute with  $\hat\gamma^{*0}$ and ${\hat\gamma}^{*12}$. As these
elements anti-commute any one of them can be chosen to provide
the third eigenvalue pair that distinguishes fermions: the following 
analysis is based on choosing $\hat\gamma^{bc}=\hat\gamma^{\pi a}$.
The eight fermions distinguished by binary eigenvalues of $\hat\gamma^{*0}$,
${\hat\gamma}^{*12}$ and $\hat\gamma^{\pi a}$ can now be identified
as fermion doublets, showing that all fermions are 
one of the eight components of a doublet.  Given 
that the Dirac equation distinguishes 
four electron states, the only other leptons that can be 
distinguished by three binary eigenvalues are the neutrinos. 
Leptons form a doublet described by the quantum numbers
A, B, C, viz.
$$
{\rm e^-\!\!} \uparrow (\bar1 1 1), \> {\rm e^-\!\!}\downarrow (1 1 1),
\> {\bar\nu\!\!}\uparrow (\bar1 \bar1 1), {\bar\nu\!\!}\downarrow (1 \bar1 1),
\>{\nu\!\!}\uparrow (\bar1 1 \bar1),\> {\nu\!\!}\downarrow (1 1 \bar1),\>
{\rm e^+\!\!} \uparrow (\bar1 \bar1 \bar1), {\rm e^+\!\!}\downarrow (1\bar1 \bar1), \eqno (2.5)
$$
with charges given by
$$
{\cal Q} = {1\over 2}(\hat{\gamma}^{*0} + i\hat{\gamma}^{\pi a}) 
= {1\over 2}(\hat{\gamma}^{\rm B} + \hat{\gamma}^{\rm C})\to \rm Q = {1\over 2}(B+C).  \eqno (2.6)
$$
Q can be interpreted as a sum of the time and parity contributions 
to the total charge. Their interaction energy is proportional to
$$
{\cal Q}^2= {1\over 2}({\bf 1}+\hat{\gamma}^{\rm B} \hat{\gamma}^{\rm C})\to \rm Q^2 = {1\over 2}(1+BC), \eqno (2.7)
$$
where $\bf 1$ denotes the unit element of $Cl_{3,3}$. This gives  $\rm Q^2=1$ for 
electrons and positrons and zero for neutrinos and anti-neutrinos,
suggesting that ${\cal Q}^2$ provides the main contribution to electron 
and positron masses. (2.6) and (2.7) show that the algebraic expression
for units of energy is $\hat{\gamma}^{*0}\hat{\gamma}^{\pi a}$.

Discrete symmetries are related to the properties of elementary fermions
by expressing them in terms of elements of $Cl_{3,3}$.
Inversion of all three spatial coordinates corresponds to  
a change in their parity, conventionally defined as 
$$
{\bf \hat P}: \hat{\gamma}^{\mu}\rightarrow \hat{\gamma}^0\hat{\gamma}^{\mu}
(\hat{\gamma}^0)^{-1} = \hat{\gamma}_{\mu}, \>{\rm or \>\>\> \bf v\to
-\bf v},                                                             \eqno (2.8)
$$
where $\hat{\gamma}^0=\hat{\gamma}_0$ is the time direction.
As the reversal of any two coordinate
directions can be achieved with a 180$^0$ rotation about the third, 
reversing any one spatial coordinate direction 
also produces a parity change.

$Cl_{3,3}$ has seven commuting elements, viz.
$\hat\gamma^{*31},\>\hat\gamma^{*2a},\>\hat\gamma^{\pi a}, 
\>\hat\gamma^{*0},\>\hat\gamma^{*0\pi a},\>\hat\gamma^{*031},\>\hat\gamma^{*02a}$.
Any three of these that are not related by multiplication
have eigenvalues that distinguish the eight leptons. 
$\hat\gamma^{*13},\>\hat\gamma^{*2a},\>\hat\gamma^{\pi a}$ have 
the interpretations in (2.4), where $\hat\gamma^{*2}$ 
is the spatial orientation and direction of the spin. 
Changes in parity are produced by two transformations, viz.
$$\eqalign{
	\hat\gamma^{*2b}\hat\gamma^{*1}\hat\gamma^{*2b}=&\hat\gamma^{*1},\>\>\>
	\hat\gamma^{*2b}\hat\gamma^{*2}\hat\gamma^{*2b}=-\hat\gamma^{*2},\>\>\>
	\hat\gamma^{*2b}\hat\gamma^{*3}\hat\gamma^{*2b}=\hat\gamma^{*3},\>\>\>
	\hat\gamma^{*2b}{\bf v}\hat\gamma^{*2b}=-{\bf v}, \cr
	\hat\gamma^{*2c}\hat\gamma^{*1}\hat\gamma^{*2c}=&\hat\gamma^{*1},\>\>\>
	\hat\gamma^{*2c}\hat\gamma^{*2}\hat\gamma^{*2c}=-\hat\gamma^{*2},\>\>\>
	\hat\gamma^{*2c}\hat\gamma^{*3}\hat\gamma^{*2c}=\hat\gamma^{*3},\>\>\> 
	\hat\gamma^{*2c}{\bf v}\hat\gamma^{*2c}=-{\bf v}, \cr }           \eqno (2.9)
$$
both of which satisfy $(\hat\gamma^{*2b})^2 = (\hat\gamma^{*2c})^2  = {\bf 1}$, so that
(2.9) describes similarity transformations. They also give
$$
\hat\gamma^{*2b}\hat\gamma^{\pi a}\hat\gamma^{*2b}=-\hat\gamma^{\pi a},\>\>\>
\hat\gamma^{*2c}\hat\gamma^{\pi a}\hat\gamma^{*2c}=-\hat\gamma^{\pi a},         \eqno (2.10)
$$
changing the sign of $\hat\gamma^{\pi a}$. As equations (2.9) and (2.10) depend only
on the multiplication properties of the $\hat\gamma$'s, they hold in any Lorentz frame, 
always producing a change in sign of a single spatial coordinate direction, 
with a consequent change in the parity of the coordinates. 
It is therefore consistent to ascribe the {\it intrinsic parity} of leptons 
to the eigenvalues of $\hat\gamma^{\pi a}=\hat\gamma^{bc}$.

Given that the eigenvalues of $\hat{\gamma}^{*0}$ determine the sign of time intervals
in the rest frame of the fermion, time reversal must be expressed in terms of
an element of the algebra that anti-commutes with $\hat{\gamma}^{*0}$ and
commutes with the unit spatial displacements 
$\hat\gamma^{*1},\>\hat\gamma^{*2},\>\hat\gamma^{*3}$. There are
two possibilities, viz. $\hat\gamma^{*0a}$ and
$\hat\gamma^{*\pi 0} = \hat\gamma^{*1}\hat\gamma^{*2}\hat\gamma^{*3}$.
These satisfy $(\hat\gamma^{*0a})^2 = -{\bf 1}$ and $(\hat\gamma^{*\pi 0})^2 ={\bf 1}$.
The most satisfactory choice is $\hat\gamma^{*\pi 0}$, which 
can be expressed as the similarity transformation
$$
\hat\gamma^{*0}\to (\hat\gamma^{*\pi 0})^{-1}\,\hat\gamma^{*0}\,\hat\gamma^{*\pi 0} 
= -\hat\gamma^{*0}.                        \eqno (2.11) 
$$
Identification of these space-time symmetries with quantum numbers ensures their 
invariance in physical processes.

\beginsection 3. Reformulation of the Dirac equation

The $Cl_{3,3}$ reformulation of the Dirac equation is 
obtained by replacing Dirac's 4$\times$4 $\gamma$-matrices with 
8$\times$8 matrix representations of the $\hat\gamma$.
Relativistic energy/momentum conservation for free 
particles is conventionally expressed as 
${\bf p}^2 = E^2 - {\vec p}^{\, 2} = m^2$, 
where $m$ is the particle mass, ${\vec p}= (p_1, p_2, p_3)$ 
its momentum 3-vector, and $E$ its energy. This corresponds 
to the Lorentz invariant relation
$$
 {\bf p} = m\hat\gamma^{*0}= p_\mu\hat\gamma^\mu, \eqno(3.1) 
$$
where $E = p_0$ and $m=p_{*0} $. The $\hat\gamma^\mu$ 
define unit measures in an observer's space-time
coordinates and $\hat\gamma^{*0}$ is the unit of 
{\it proper time} measured in the coordinate frame at rest 
with respect to the particle. 
Equation (3.1) can be written $\>\>p_\mu\hat\gamma^\mu - m\hat\gamma^{*0} =0\>$, 
which is the classical relativistic form of the Dirac equation. 
The factor $\hat\gamma^{*0}$ ensures that $m$ is always positive. 
  
Equation (3.1) requires adaptation to provide
an explicit statement of lepton energy conservation. 
In this case $\bf p$ spans all eight states 
listed in (2.5) which, following (2.7), gives   
$$
p_{*0} = m_{\rm e}{1\over 2}{\rm (1+BC)} + m_\nu{1\over 2}\rm (1-BC),        \eqno(3.2)
$$ 
where $m_e$ is the mass of electrons and positrons and $m_\nu$ is the
mass of neutrinos and anti-neutrinos. Hence the representation of 
the energy momentum 4-vector in terms of lepton rest frame coordinates 
is given by
$$
{\bf p}=\hat\gamma^{*0}p_{*0} = diag(m_e, m_e, m_\nu, m_\nu, m_\nu, m_\nu, m_e, m_e),\eqno(3.3)
$$
where the eight lepton states are labelled in the order in (2.5). 
In observer's coordinates the $\hat\gamma$ representation of ${\bf p}$ 
has the block diagonal form given in Appendix B, viz.
$$
{\bf p} \equiv \hat\gamma^{\mu} p_\mu 
= \left(\matrix{ {\bf p}_+ &0 \cr
	0 & {\bf p}_-\cr} \right).                                              \eqno (3.4)
$$
Here the $\pm$ labels refer to the parity of the coordinates, defined
by the ${\rm C}=\pm 1$ eigenvalues of $i\hat\gamma^{\rm C}$. 
The columns in $\bf p$ correspond to the eight lepton states 
labelled in the same order as in (3.3). 
Following the matrix representation in Appendix B, 
and expressing the parity difference as 
a change in sign of the $x_2$ coordinate, 
$$
{\bf p}_+  =
\left(\matrix{p_0       &0          & p_2       &-p_1-ip_3\cr
	0       &p_0        &-p_1+ip_3  &-p_2\cr
	-p_2    &p_1+ip_3   &-p_0       &0\cr
	p_1-ip_3&p_2        &0          &-p_0\cr}
\right),                                                                   \eqno(3.5+)
$$  
$$                                                     
{\bf p}_-  =
\left(\matrix{p_0         &0          & -p_2       &-p_1-ip_3\cr
	0         &p_0        &-p_1+ip_3 &p_2\cr
	p_2        &p_1+ip_3 &-p_0        &0\cr 
	p_1-ip_3&-p_2       &0           &-p_0\cr}
\right).                                                                   \eqno(3.5-)
$$
Both of these expressions are diagonal in the lepton rest frame, with
the $p_0\to p_{*0}$ values given in (3.3). Transformations relating
coordinate frame have the form ${\bf p}\to {\bf L p L^{-1}}$, 
where the block diagonal matrix representation of the Lorentz transformation 
$\bf L$ is given in equation $(B.3)$ of Appendix B.

The columns in the matrix representation of $\bf p$ can be
identified as spinors defined by the lepton 
states in the order given in (2.5). The first and last two 
columns have $p_{*0}= m_e$ and correspond to electron states.
The first two columns have the same 
form as the $u_i(p)$ factors in the free electron spinor solutions
of the Dirac equation given in (4.51) of [18]. Similarly, 
the last two have the same form as the $v_i(p)$ factors in the free 
positron spinor solutions of the Dirac equation given 
in  (4.52) of [18]. The only differences in both cases 
are normalization and a change of sign in $p_1$ and $p_2$
that corresponds to a $\pi$ rotation in the x-y plane.
This show the structure of free fermion spinors to be
determined by the algebra. 

The standard procedure for obtaining wave-equations 
from their classical counterparts is to
replace ${\vec p}\>$ by the operator 
$-i\nabla = -i(\partial_1, \partial_2,\partial_3)$ and 
$E$ by the operator $ i\partial_0$, so that
$$
{\bf p} =\hat\gamma_\mu p^\mu \to  i{\bf D}= i\hat\gamma^{*0}\partial_{*0}
= i\hat\gamma^\mu \partial_\mu.                                           \eqno (3.6)
$$
The wave equation is produced by the action of $i{\bf D}$ 
on a space-time dependent wave function $\phi(x^\mu)$, giving the 
relativistic free particle wave equation
$$
(i{\bf D} - p_{*0}{\hat\gamma}^{*0})\phi(x^\mu)=0,                         \eqno (3.7)
$$ 
where $\phi(x^\mu)$ describes the wave motion of any lepton.
(3.7) differs from the Dirac equation only in the presence
of the factor ${\hat\gamma}^{*0}$, which ensures that positron 
masses are positive. $\phi$ can be expressed in terms 
of observer's or lepton coordinates, viz.
$$
\phi =  \exp (-ip_\mu x^\mu) =  \exp (-ip_{*0} x^{*0}),                                   \eqno (3.8)      
$$
where $x^{*0} $ is the proper time, i.e. time measured in the 
lepton coordinate frame. In observer's coordinates
$$
i{\bf D}\phi = i\hat\gamma^\mu \partial_\mu \exp (-ip_\mu x^\mu)
=\hat\gamma^\mu p_\mu \phi={\bf p}\phi .                                   \eqno (3.9a)      
$$
In lepton coordinates,
$$
i{\bf D}\phi = i\hat\gamma^{*0} \partial_{*0} \exp (-ip_{*0} x^{*0}) 
=\hat\gamma^{*0} p_{*0}\phi= {\bf p}\phi                    \eqno (3.9b)      
$$
where $p_{*0}$ is given in (3.3). Hence the free lepton Dirac equation becomes 
$$
(i{\bf D}-{\bf p})\phi= 0 .                  \eqno (3.10)      
$$

This equation can be modified to include interactions
with electromagnetic fields simply by adding the field momentum contribution
to the particle momentum, as is done in Lagrangian
theory. For example, the electromagnetic
contribution $e \cal Q\bf A$ can be added to the free particle momentum $\bf p$
to produce the generalized, or {\it canonical}, momentum
$$
{\bf p'}= {\bf p} + e \cal Q\bf A .                                 \eqno (3.11)
$$
With this modification, (3.10) becomes
$$
 i{\bf D}\phi = i{\hat\gamma}^\mu \partial_\mu \phi  
=({\bf p}+ e {\bf A}{\cal Q})\phi.                                   \eqno(3.12) 
$$
The factor ${\bf p}+ e {\bf A}{\cal Q}$ can be brought down 
from the exponent if 
$$
{\phi }= \>\exp(i\int{ p'}_\mu dx^\mu ),       \eqno (3.13)
$$
with
$$
{\bf p'}  = \>i\hat\gamma^{\mu} p'_\mu = {\bf p}  +e{\bf A}{\cal Q},\>\>
{\rm where\>\>} {\bf A} =\> A_\mu \hat\gamma^{\mu},\>\>\>\>
{\rm and \>\>}{\bf A}{\cal Q}=\>-{1\over 2}A_\mu \hat\gamma^{\mu}(\hat\gamma^{*0}+i\hat\gamma^{\pi a} ).\eqno (3.14)
$$

\beginsection \S4. Reformulation of the electro-weak interaction

It was shown in \S3 that the $Cl_{3,3}$ algebra provides a 
description of all eight states in the lepton doublet. It
also contains a description of the weak interaction 
between the components of this doublet.
This is related to the Standard Model 
formulation (e.g. \S15.3 of [18]) by replacing the Pauli matrix unit
vectors $\sigma_i$ by anti-commuting time-like 
elements $\hat\gamma^{(i)} $ of $Cl_{3,3}$.
Their identification is guided by expressing the  
$\sigma_i$ in the matrix notation employed in Appendix B, viz. 
$\sigma_1= {\bf Q}, \>\>\sigma_2 = i{\bf P},\>\> \sigma_3 = {\bf R}$.

As $\hat\gamma^{\pi a}$ takes eigenvalues for all lepton states
it must correspond to the diagonal Pauli matrix $\sigma_3 = {\bf R}$ which,
according to the table in Appendix B, is given by
$$
\hat\gamma^{(3)} \equiv i\hat\gamma^{\pi a}= i\hat\gamma^{bc} 
= -{\bf I}\otimes {\bf I}\otimes{\bf R} = \hat\gamma^{\rm C}.  \eqno (4.1)
$$
$\hat\gamma^{(1)},\>\hat\gamma^{(2)}$ can now be identified as 
$$
\gamma^{(1)} \equiv \hat\gamma^{*2c}= {\bf I}\otimes {\bf I}\otimes{\bf Q},
\>\> \gamma^{(2)}\equiv -i\hat\gamma^{*2b}= {\bf I}\otimes {\bf I}\otimes{\bf P}.      \eqno (4.2)
$$
The factor $\hat\gamma^{*2}$ is the spatial orientation of the spin
described by $\hat\gamma^{31}$.
This breaks the Coleman-Mandula condition that the  
$\hat\gamma^{(i)}$ should commute with the matrices that define the space-time 
coordinates. Nevertheless, overall Lorentz invariance holds 
as spatial rotations retain the relationship between the expressions
for the $\hat\gamma^{(i)}$ and the spin orientation.

Raising and lowering operators that describe the weak bosons are 
$$\eqalign{
	\hat\gamma^- = \>&\>{1\over  2}\,(\hat\gamma^{(1)} + \hat\gamma^{(2)}) 
	= \>\>\>{1\over 2}{\bf I}\otimes {\bf I}\otimes ({\bf P}+ {\bf Q})
	=\left(\matrix{ 0&0\cr {\bf I}\otimes {\bf I}&0\cr }\right),\cr
	\hat\gamma^+ = \>&\>{1\over 2}\,(\hat\gamma^{(2)}-\hat\gamma^{(1)})
	= \>\>\>{1\over 2}{\bf I}\otimes {\bf I}\otimes ({\bf Q}-{\bf P})
	=\left(\matrix{ 0& {\bf I}\otimes {\bf I}\cr 0&0\cr }\right),\cr }                 \eqno (4.3)
$$
and satisfy
$$\eqalign{\hat\gamma^- \hat\gamma^- = 0,\>&\> \hat\gamma^+ \hat\gamma^+ = 0,\cr
	\hat\gamma^-\hat\gamma^+ + \hat\gamma^+\hat\gamma^- = {\bf 1}_3,\>&
	\>\hat\gamma^+\hat\gamma^- -\hat\gamma^- \hat\gamma^+ = \hat\gamma^{(3)}.}         \eqno (4.4)
$$
The matrix expression makes it clear that $\hat\gamma^+$ has the effect of 
changing C=1 to C=$1$, and  $\hat\gamma^-$ has the effect of 
changing C=1 to C=$-1$. 

Defining 
$ W^+_\kappa = W^1_\kappa - iW^2_\kappa $ and $W^-_\kappa = W^1_\kappa +iW^2_\kappa $, 
the weak potential can be expressed as
$$
{\bf W} = \hat\gamma^\kappa{\bf W_\kappa}=\hat\gamma^\kappa({\bf W}_\kappa^+\hat\gamma^{+}+ {\bf W}_\kappa^- \hat\gamma^{-} +{\bf W}^3_\kappa \hat\gamma^{(3)}),                                                       \eqno(4.5)
$$
giving the weak interaction
$$
{\bf X_\kappa} =  {g_W\over 2}{\bf W}_\kappa = {g_W\over 2} ({\bf W}_\kappa^+\hat\gamma^{+}+ {\bf W}_\kappa^- \hat\gamma^{-} +{\bf W}^3_\kappa \hat\gamma^{(3)}).                                                    \eqno (4.6)
$$
\vskip5pt
The action of $\hat\gamma^\pm $ on ${\bf p} $ gives
$$\hat\gamma^+ {\bf p} =\left(\matrix{ 0&0\cr \>{\bf p}_+&0 \cr} \right),\>\>
\hat\gamma^-{\bf p}=\left(\matrix{ 0&\>{\bf p}_-\cr 0&0 \cr} \right).                 \eqno (4.7)
$$
The physical interpretation of these equations is that the positively charged 
boson $\hat\gamma^+$ adds a charge to fermions with negative or zero charges,
for example converting $e^-\rightarrow  \nu$ and $\bar\nu\rightarrow  e^+ $.
Similarly $\hat\gamma^-$ subtracts a charge, so that $e^+ \rightarrow \bar\nu$
and $\nu\rightarrow  e^- $. It follows that both $\hat\gamma^+$ and $\hat\gamma^-$
change the parity of the coordinate frame, in agreement with the observed parity 
change produced by the weak interaction.

Electro-weak unification and the role of the Higgs boson are not discussed here as
they are not directly relevant to the theme of this work.

\beginsection \S5. Physical interpretation of $Cl_{5,5}$

The ten anti-commuting generators $\hat\gamma^{\,i}$ of the 
algebra $Cl_{5,5}$ are constructed by adding the two space-like
generators $\hat\gamma^{\,4}$, $\hat\gamma^{\,5}$ and the two 
time-like generators $\hat\gamma^{\,d}$, $\hat\gamma^{\,e}$ 
to the six generators of its $Cl_{3,3}$ sub-algebra.
Following (A.1), the time direction is specified by the sign
of the ten dimensional volume element of $Cl_{5,5}$ which, 
in fermion rest frames, is
$$
\hat\gamma^{\,*0} =\hat\gamma^{\,*1}\hat\gamma^{\,*2}\hat\gamma^{\,*3}\hat\gamma^{\,4}
\hat\gamma^{\,5}\hat\gamma^{\,a}\hat\gamma^{\,b}\hat\gamma^{\,c}\hat\gamma^{\,d}\hat\gamma^{\,e}
=-\hat\gamma^{\,13}\hat\gamma^{\,2a}\hat\gamma^{\,bc}\gamma^{\,4}\hat\gamma^{\,5}
\hat\gamma^{\,d}\hat\gamma^{\,e}.                                      \eqno (5.1)
$$                                             
$\hat\gamma^4$ and $\hat\gamma^5$ are not observed spatial dimensions, 
so the (3.2) expression $\hat\gamma^\pi$ for unit space-time volumes
is unchanged. Five commuting elements of $Cl_{5,5}=Cl_{3,3}\otimes Cl_{2,2}$ 
define the five binary quantum numbers A,B,C,D,E. The sub-algebra $Cl_{3,3}$ 
defined A,B,C, which were shown in \S3 to distinguish the eight lepton 
states in a single generation. More generally, they will be found to
distinguish the eight states of all fermion doublets. 

The quantum numbers D and E are defined by the two commuting time-like elements, 
viz. $\hat\gamma^{\rm D}=\hat\gamma^{4d}$ and $\hat\gamma^{\rm E}=\hat\gamma^{5c}$.
All five quantum numbers distinguish $4\times 8 = 32$ states, comprising four
doublets, one of which is the lepton doublet. Fermions have B=1, so that  
$\hat\gamma^{\rm B}$ can be expressed as the truncated matrix 
$\hat\gamma^0 \equiv diag(1 1 1 1)$ and the elements 
$\hat\gamma^{\rm D},\>\hat\gamma^{\rm E},\>-\hat\gamma^{\rm BDE}$ 
have the diagonal representations
$$	\hat\gamma^{\rm D}\>\equiv diag(\bar1 1 1 \bar1),\>\>\>
	\hat\gamma^{\rm E} \>\equiv diag( \bar1 1 \bar1 1),\>\>\>
	-\hat\gamma^{\rm DEB}= -\hat\gamma^{\rm D}\hat\gamma^{\rm E}\hat\gamma^{\rm B} 
	\equiv diag(1 1 \bar 1\bar 1).                                    \eqno (5.2)
$$
Elementary fermions correspond to diagonal matrices with a single non-zero
element. Leptons and quark colours are related to the quantum numbers B,D,E as follows
$$\eqalign{
{\bf l}\equiv & {1\over 4}{\rm(\hat\gamma^B -\hat\gamma^D -\hat\gamma^E +\hat\gamma^{DEB})}
	= diag(0 0 0 1 ),\cr
{\bf q}_b =&{1\over 4}{\rm(\hat\gamma^B +\hat\gamma^D +\hat\gamma^E +\hat\gamma^{DEB})}
	\equiv diag(1 0 0 0),\cr
{\bf q}_g =& {1\over 4}{\rm(\hat\gamma^B +\hat\gamma^D -\hat\gamma^E -\hat\gamma^{DEB})}
	\equiv diag(0 1 0 0),\cr
{\bf q}_r=&{1\over 4}{\rm(\hat\gamma^B -\hat\gamma^D +\hat\gamma^E -\hat\gamma^{DEB})}
	\equiv diag(0 0 1 0).}                                                     \eqno (5.3)
$$
Quark colour labels $b,g,r$ are not distinguished by their physical properties,
so that their relation to D,E descriptions is arbitrary.

The quarks can be d or u, and the leptons e$^-$ or $\nu$, 
according to the sign of C. Anti-fermion descriptions are obtained by reversing the
signs of all five quantum numbers. Equation (5.3) is consistent with the $\rm Q_B$ 
contribution to fermion charges being 
$$
\rm Q_B ={1\over 6}( D + E - BDE ) = {1\over 6}\>{\rm for\> quarks,\>}
 -{1\over 2}\>{\rm for\> leptons},                                                  \eqno(5.4)
$$ 
giving electric charge operator, which generalises (2.6), as 
$$
{\cal Q} = \rm {1\over 6}(\hat\gamma^D + \hat\gamma^E-\hat\gamma^{BDE}) - {1\over 2} \hat\gamma^C. \eqno (5.5)
$$
 
Quark/quark interactions mediated by gluons produce colour exchange, and
are conventionally expressed in terms of the eight generators of an 
irreducible representation of SU(3). References [4-8] have 
already pointed out that these generators can be expressed in terms of 
the elements of Clifford algebras. No individual quarks or gluons 
have been observed, showing them to be confined to regions of space 
inside hadrons and glueballs. 

\vfill\eject

\beginsection \S6. Physical interpretation of $Cl_{7,7}$ 

The fourteen generators of $Cl_{7,7}$ are denoted $\hat\gamma^i,\> \{i= 1-7,\> a-g\}$.
Their product defines unit the volumes 
$\hat\gamma^0=\hat\gamma^1\hat\gamma^{2}...\hat\gamma^{f}\hat\gamma^{g}$
in its fourteen dimensional space, which are
are identified as unit time intervals. This is 
consistent with the unit time intervals identified in the
$Cl_{3,3}$ and $Cl_{5,5}$ sub-algebras.
The (2.2) expression $\hat\gamma^\pi$ for unit space-time volumes
is retained. Quantum numbers A,B,C,D,E were identified 
in the analysis of the sub-algebra $Cl_{5,5}$. Two additional
quantum numbers F,G, which are related to the generators 
$\hat\gamma^6,\>\hat\gamma^f,\>\hat\gamma^7,\>\hat\gamma^g$,
distinguish fermion generations. The following analysis follows
the same pattern as that used to distinguish leptons and quarks, with 
correspondences F$\leftrightarrow$D, G$\leftrightarrow$E.

F, G are eigenvalues of the commuting matrices
$$
\hat\gamma^{\rm F} = \hat\gamma^{6f} = \hat\gamma^6\hat\gamma^f,\>\>\>
\>\hat\gamma^{\rm G} = \hat\gamma^{7g} = \hat\gamma^7\hat\gamma^g, \eqno (6.1)
$$
which have the diagonal representations
$$	\hat\gamma^{\rm F}\>\equiv diag(\bar1 1 1 \bar1),\>\>\>
\hat\gamma^{\rm G} \>\equiv diag( \bar1 1 \bar1 1),\>\>\>
-\hat\gamma^{\rm BFG}= -\hat\gamma^{\rm B}\hat\gamma^{\rm F}\hat\gamma^{\rm G} 
\equiv diag(1 1 \bar 1\bar 1).                                               \eqno (6.2)
$$
These define the four generations
$$\eqalign{
	{\bf g}_1 =&{1\over 4}{\rm(\hat\gamma^B +\hat\gamma^F +\hat\gamma^G +\hat\gamma^{BFG})}
	\equiv diag(1 0 0 0),\cr
	{\bf g}_2 =& {1\over 4}{\rm(\hat\gamma^B +\hat\gamma^F -\hat\gamma^G -\hat\gamma^{BFG})}
	\equiv diag(0 1 0 0),\cr
	{\bf g}_3=&{1\over 4}{\rm(\hat\gamma^B -\hat\gamma^F +\hat\gamma^G -\hat\gamma^{BFG})}
	\equiv diag(0 0 1 0),\cr
	{\bf g}_4 =& {1\over 4}{\rm(\hat\gamma^B -\hat\gamma^F -\hat\gamma^G +\hat\gamma^{BFG})}
	\equiv diag(0 0 0 1 ).}                                                     \eqno (6.3) 
$$
As fermions in the first three generations carry the same charge, the assignment of 
${\bf g}_1$, ${\bf g}_2$, ${\bf g}_3$ to specific F,G labels is arbitrary.

The $\rm Q_C$ contribution to fermion electric charges is $-{1\over 2}$( F +G - BFG)BC,
giving total charges
$$
\rm Q = Q_B + Q_C ={1\over 6}( D + E- BDE) - {1\over 2}( F + G - BFG )BC,        \eqno (6.4)
$$
with the corresponding charge operator
$$
{\cal Q} =\rm {1\over 6}( \hat\gamma^D +\hat\gamma^E-\hat\gamma^B\hat\gamma^D\hat\gamma^E)
- {1\over 2}( \hat\gamma^F+ \hat\gamma^G- \hat\gamma^B\hat\gamma^F\hat\gamma^G )\hat\gamma^B \hat\gamma^C.                      \eqno (6.5)
$$

Observed mesons interchange generations 1 to 3, but do not interact with 4th
generation fermions. This is analogous with the fact that gluons 
do not interact with leptons, suggesting that 4th generation fermions
are spatially separated from fermions in the first three generations. 
Equation (6.4) gives the same charges on fermions in all three known generations,
but predicts very different charges on fermions in the currently unobserved
4th generation. The existence of G4 fermions is doubted by most particle 
physicists because of the clear experimental evidence quoted,
for example, in [19], p.344, Fig.9.30, that only three generations 
of neutrinos exist. However, as the predicted G4 leptons do not include neutrinos,
this interpretation of the experimental evidence 
is invalid. G4 fermions have not been observed, inevitably suggesting 
that their composites could be the constituents of dark matter and/or
dark energy. 

\vfill\eject

\beginsection \S7. Summary of results

A seven quantum number description of all the elementary 
fermions has been determined. Quantum numbers B,C,D,E,F,G 
distinguish the elementary fermions, determine 
their electric charges and provide their {\it signatures}. 
Spin directions are described by $\rm A=\pm1$. 
All the CDEFG fermion signatures with B=1 are given below. 
Corresponding anti-fermions have opposite signs of all 
seven quantum numbers.

$$\vbox
{\settabs 11 \columns
	\+&{\bf Definitive (B=1) signatures for all four generations of fermions}\cr 
	\+&|||||||||||||||||||||||||||||||||||||||||\cr
	\+&quark&C&D&E&F&G$\>\>\>\>\>\>\>\rm gene$&$\rm ration$ & $\rm Q_B$&$\rm Q_C$&Q\cr
	\+&|||||||||||||||||||||||||||||||||||||||||\cr
	\+&${\rm u}_b$    &$-1$   &$\>\>1$&$\>\>1$&$\>\>1$ &$\>\>1$   &[1+]&1/6&1/2&$\>\>2/3$ \cr
	\+&${\rm u}_r$    &$-1$   &$\>\>1$&$-1$   &$\>\>1$ &$\>\>1$   &[1+]&1/6&1/2&$\>\>2/3$ \cr
	\+&${\rm u}_g$    &$-1$   &$-1$   &$\>\>1$&$\>\>1$ &$\>\>1$   &[1+]&1/6&1/2&$\>\>2/3$ \cr 
	\+ &&&\cr
	\+&${\rm d}_b$    &$\>\>1$&$\>\>1$&$\>\>1$&$\>\>1$ &$\>\>1$&$[1-]$&1/6&$-1/2$&$-1/3$  \cr
	\+& ${\rm d}_r$   &$\>\>1$&$\>\>1$&$-1$   &$\>\>1$ &$\>\>1$&$[1-]$&1/6&$-1/2$&$-1/3$  \cr
	\+&${\rm d}_g$    &$\>\>1$&$-1$   &$\>\>1$&$\>\>1$ &$\>\>1$&$[1-]$&1/6&$-1/2$&$-1/3$  \cr 
	\+&|||||||||||||||||||||||||||||||||||||||||\cr
	\+&${\rm c}_b $       &$-1$   &$\>\>1$&$\>\>1$&$\>\>1$    &$-1$   &[2+]&1/6&$1/2$&$\>\>2/3$\cr
	\+&${\rm c}_r $       &$-1$   &$\>\>1$&$-1$   &$\>\>1$    &$-1$   &[2+]&1/6&$1/2$&$\>\>2/3$\cr
	\+& ${\rm c}_g $      &$-1$   &$-1$  &$\>\>1$ &$\>\>1$    &$-1$   &[2+]&1/6&$1/2$&$\>\>2/3$\cr 
	\+ &&&\cr
	\+& ${\rm s}_b $      &$\>\>1$&$\>\>1$&$\>\>1$&$\>\>1$    &$-1$   &$[2-]$&1/6&$-1/2$& $-1/3$  \cr
	\+&${\rm s}_r $       &$\>\>1$&$\>\>1$&$-1$   &$\>\>1$    &$-1$   &$[2-]$&1/6&$-1/2$& $-1/3$ \cr
	\+& ${\rm s}_g $      &$\>\>1$&$-1$   &$\>\>1$&$\>\>1$    &$-1$   &$[2-]$&1/6&$-1/2$& $-1/3$  \cr
	\+&|||||||||||||||||||||||||||||||||||||||||\cr		
	\+ &${\rm t}_b $  &$-1$   &$\>\>1$&$\>\>1$&$-1$  &$\>\>1$   &[3+]&1/6&$1/2$&$\>\>2/3$ \cr
	\+ &${\rm t}_r $  &$-1$   &$\>\>1$&$-1$   &$-1$  &$\>\>1$   &[3+]&1/6&$1/2$&$\>\>2/3$ \cr
	\+ &${\rm t}_g $  &$-1$   &$-1$   &$\>\>1$&$-1$  &$\>\>1$   &[3+]&1/6&$1/2$&$\>\>2/3$\cr 
	\+ &&&\cr
	\+ &${\rm b}_b $  &$\>\>1$&$\>\>1$&$\>\>1$&$-1$  &$\>\>1$   &$[3-]$&1/6&$-1/2$&$-1/3$    \cr	
	\+ &${\rm b}_r $  &$\>\>1$&$\>\>1$&$-1$   &$-1$  &$\>\>1$   &$[3-]$&1/6&$-1/2$&$-1/3$   \cr	
	\+ &${\rm b}_g $  &$\>\>1$&$-1$   &$\>\>1$&$-1$  &$\>\>1$   &$[3-]$&1/6&$-1/2$&$-1/3$  \cr
	\+&|||||||||||||||||||||||||||||||||||||||||\cr
	\+&$q_b(-4/3)$      &$-1$   &$\>\>1$   &$\>\>1$  &$-1$&$-1$ &$[4-]$&1/6&$-3/2$&$-4/3$\cr
	\+&$q_r(-4/3)$      &$-1$   &$\>\>1$   &$-1$     &$-1$&$-1$ &$[4-]$&1/6&$-3/2$&$-4/3$\cr
	\+&$q_g(-4/3)$      &$-1$   &$-1$      &$\>\>1$  &$-1$&$-1$ &$[4-]$&1/6&$-3/2$&$-4/3$\cr
	\+ &&&\cr
	\+&$q_b(5/3)$      &$\>\>1$&$\>\>1$    &$\>\>1$   &$-1$&$-1$&[4+]  &1/6&3/2&$\>\>5/3$\cr 
	\+&$q_r(5/3)$      &$\>\>1$&$\>\>1$    &$-1$      &$-1$&$-1$&[4+]  &1/6&3/2&$\>\>5/3$\cr 
	\+&$q_g(5/3)$      &$\>\>1$&$-1$       &$\>\>1$   &$-1$&$-1$&[4+   &1/6&3/2&$\>\>5/3$\cr 
	\+&|||||||||||||||||||||||||||||||||||||||||\cr	  	
	\+ &$\nu_e $       &$-1$&$-1$   &$-1$   &$\>\>1$ &$\>\>1$ &[1+]  &$-1/2$&1/2&$\>\>0$\cr	
	\+ &$\nu_\mu $     &$-1$&$-1$   &$-1$   &$\>\>1$ &$-1$    &[2+]  &$-1/2$&1/2&$\>\>0$\cr	
	\+ &$\nu_\tau $    &$-1$&$-1$   &$-1$   &$-1$    &$\>\>1$ &[3+]  &$-1/2$&1/2&$\>\>0$\cr
	\+ &$l(-2)$        &$-1$&$-1$   &$-1$   &$-1$    &$-1$    &[4+]  &$-1/2$&$-3/2$&$-2    $\cr
	\+ &&&\cr
	\+ &e$^-$       &$\>\>1$&$-1$   &$-1$   &$\>\>1$ &$\>\>1$&$[1-]$ &$-1/2$&$-1/2$&$-1$    \cr
	\+ &$\mu^-$     &$\>\>1$&$-1$   &$-1$   &$\>\>1$ &$-1   $&$[2-]$ &$-1/2$&$-1/2$&$-1$    \cr
	\+ &$\tau^-$    &$\>\>1$&$-1$   &$-1$   &$-1$    &$\>\>1$&$[3-]$ &$-1/2$&$-1/2$&$-1$    \cr 
	\+ &$l(1)$      &$\>\>1$&$-1$   &$-1$   &$-1$    &$-1$   &$[4-]$ &$-1/2$&$\>\>3/2$&$\>\>1$  \cr	
	\+&|||||||||||||||||||||||||||||||||||||||||\cr}
$$

\beginsection \S8. Quantum number conservation 

Each of the $2^7$ distinct elementary fermions is specified 
by a unique combination of the seven binary quantum numbers A-G, which 
have the physical interpretations defined in earlier sections. 
Experiment shows that, although fermions can change identity in decays 
and interactions, their quantum number descriptions satisfy the \vskip 2pt
{\bf Conservation Law:} {\it All seven quantum numbers are conserved in fermion decays and interactions.}
\vskip5pt
It is convenient to employ the condensed notation introduced in \S2 by
writing $\bar1$ instead of $-1$. Fermions of either 
spin are described by a {\it signature} consisting of the six 
quantum numbers  B,C,D,E,F,G. For example electrons have the signature
$\{{\rm e}^-:1\>1\>\bar1\>\bar1\>1\>1\>\} $. 
$\beta$ decay is produced by the first generation process 
$ {\rm d}\to {\rm u}+ {\rm e}^- + \bar\nu_e$ which, in terms
of signatures, is
$$ 
\{{\rm d}_b:1\>1\>1\>1\>1\>1\}=\{{\rm u}_b:1\>\bar1\>1\>1\>1\>1\} 
+\{{\rm e}^-:1\>1\>\bar1\>\bar1\>1\>1\}+\{\bar\nu_e:\bar1\>1\>1\>1\>\bar1\>\bar1\}.    \eqno (8.1)
$$
Conservation is shown by each of the six quantum numbers
on the left hand side of this equation equalling the sum of
each of the corresponding numbers on its right hand side. 
Similar examples are provided by the lepton decay
products $\mu^- \to {\rm e}^- \bar \nu_e\,\nu_\mu$ 
and $\tau^- \to {\rm e}^- \bar \nu_e\,\nu_\tau$, viz.
$$
\{\mu^-:1\>1\>\bar1\>\bar1\>1\>1\}=
\{{\rm e}^-:1\>1\>\bar1\>\bar1\>1\>\bar1\}+\{\bar\nu_e:\bar1\>1\>1\>1\>\bar1\>\bar1\}
+\{\nu_\mu:1\>\bar1\>\bar1\>\bar1\>1\>\bar1\},                              \eqno(8.2a)
$$
$$
\{\tau^-:1\>1\>\bar1\>\bar1\>\bar1\>1\}=
\{{\rm e}^-:1\>\bar1\>\bar1\>\bar1\>\bar1\>1\}+\{\bar\nu_e:\bar1\>1\>1\>1\>1\>\bar1\}
+\{\nu_\tau:1\>\bar1\>\bar1\>\bar1\>\bar1\>1\}.                               \eqno(8.2b)
$$
Possible processes can also be identified by moving a signature
on one side of these equations to the other side, and reversing 
the sign of all six quantum numbers. For example, (8.1) becomes
$$ 
\{\rm d_b:1\>1\>1\>1\>1\>1\}+\{\nu_e:1\>\bar1\>\bar1\>\bar1\>1\>1\}
=\{{\rm u}_b:1\>\bar1\>1\>1\>1\>1\}+\{{\rm e}^-:1\>1\>\bar1\>\bar1\>1\>1\},           \eqno (8.3)
$$
which describes the scattering of the $\nu_e$ neutrino by a d quark to produce
a u quark and an electron.

Fermion interactions that involve bosons do not affect 
fermion quantum number conservation. The 
established model of eight gluons that interchange the colours of a 
quark and and anti-quark is of particular interest. For example, 
using abbreviated BCDE signatures for first generation quarks,
$$
\rm \{u_r:1\bar1 1 \bar1\}+\{\bar d_g:\bar1 \bar1 1 \bar1\})
=\rm \{g_{rg}:0 0 4 \bar4\}+\{u_g:1\bar1 \bar1 1\}+\{\bar d_r:\bar1 \bar1 \bar1 1\}. \eqno (8.4)
$$
The same equation holds if the gluon signature $\rm g_{rg}$ is 
transferred to the left hand side with a change in sign of all its quantum numbers, 
so that $\rm g_{gr}= -\rm g_{rg}$. Quantum number conservation refers to individual nodes in
Feynman diagrams. Lines between nodes can correspond to either fermions or bosons. In
both cases their contribution to one node is the anti-particle of their contribution 
to the other node. Consequently interior lines in arbitrarily complex Feynman diagrams
make do not affect the quantum number conservation of their exterior fermion lines.  

Weak boson interactions $\hat\gamma^{\pm}$ change
the charges on fermions by one unit. Given the convention 
that C$=\pm 3$ fermions do not exist their description in 
terms of quantum number addition is
$$
\{\hat\gamma^+ :0\>\bar 2\>0\>0\>0\>0\},\>\>\>\>\>\>
\{\hat\gamma^- :0\> 2\>0\>0\>0\>0\>\} ,                                      \eqno (8.5)
$$
showing that bosons can also be, at least partially, described in terms of 
quantum numbers. Examples of weak bosons interactions are
$$\eqalign{
     \{\hat\gamma^- :0\>\bar 2\>0\>0\>0\>0\}+ \{\bar\nu_e:\bar1\>1\>1\>1\>\bar1\>\bar1\}=& 
    \{{\rm e}^+:\bar1\>\bar1\>1\>1\>\bar1\>\bar1\},  \cr                                
	\{\hat\gamma^+ :0\> 2\>0\>0\>0\>0\>\} +\{\nu_e:1\>\bar1\>\bar1\>\bar1\>1\>1\}  = &
	\{{\rm e}^-:1\>1\>\bar1\>\bar1\>1\>1\}.   }                                         \eqno (8.6)
$$
The action of $\{\hat\gamma^+ :0\>\bar 2\>0\>0\>0\>0\>\}$ 
on $\{{\rm e}^+:\bar1\>\bar1\>1\>1\>\bar1\>\bar1\}$ does not produce a fermion.

The description of neutrino/fermion interactions is relevant to current
experimental work. The equalities
$$
\{{\rm e}^-:1\>1\>\bar1\>\bar1\>1\>1\}+ \{{\rm \bar\nu_e}:\bar1\>1\>1\>1\>\bar1\>\bar1\}
=\{{ \mu}^-:1\>1\>\bar1\>\bar1\>1\>\bar1\}+ \{{ \bar\nu_\mu}:\bar1\>1\>1\>1\>\bar1\>1\}
=\{{\tau}^-:1\>1\>\bar1\>\bar1\>\bar1\>1\}+ \{{ \bar\nu_\tau}:\bar1\>1\>1\>1\>1\>\bar1\}  \eqno (8.7)
$$
show the following neutrino generation changing interactions to be possible:
$$
{\rm \tau}^- + {\rm \nu_e}\to {\rm e}^- + {\nu_\tau},
\>\>\>\>\>{\tau}^- +{\nu_\mu}\to{\mu}^- +{\nu_\tau},\>\>\>\>\>
 {\mu}^- + {\rm \nu_e}\to {\rm e}^- + {\nu_\mu}.                                    \eqno (8.8)
$$
Analogous interactions with hadrons are also of interest.

Baryons, composed of three quarks, can be described by the sum of the
signatures of their components, for example, $\Delta$(uds) baryons, 
protons p(uud) and neutrons n(udd) have signatures in which the quarks have
different colour labels. For example 
$$ \eqalign{\{\Lambda:3\>1\>1\>1\>3\>1\}=& \rm\{u_b:1\>\bar1\>1\>1\>1\>1\}
	+\{s_r:1\>1\>1\>\bar1\>1\>\bar1\} + \{d_g:1\>1\>\bar1\>1\>1\>1\},\cr 
	\{\rm p:3\>\bar1\>1\>1\>3\>3\}=&\rm \{u_b:1\>\bar1\>1\>1\>1\>1\}
	+\{ u_r:1\>\bar1\>1\>\bar1\> 1\> 1\} + \{d_g:1\> 1\>\bar1\>1\>1\>1\},\cr 
	\{\rm n:3\>1\>1\>1\>3\>3\}=& \rm\{u_b:1\>\bar1\>1\>1\>1\>1\}
	+\{d_r:1\>1\>1\>\bar1\>1\>1\} + \{d_g:1\>1\>\bar1\>1\>1\>1\}.}                  \eqno (8.9)
$$
The B=3 quantum numbers refer to the number of quarks 
that make up the nucleon.
The C, D and E quantum numbers in the proton signature are the 
same as those of the positron. The sum of
proton and electron signatures gives the hydrogen atom signature $\{H:4\>0\>0\>0\>4\>4\}
=\{{\rm e}^-:1\>1\>\bar1\>\bar1\>1\>1\}+\{\rm p:3\>\bar1\>1\>1\>3\>3\}$. 
Signatures for other atoms and ions can be constructed in the same way.
Quantum number conservation provides stronger constraints than charge on possible meson decays.
Mesons are conventionally distinguished by their quark structures, such as
$$\eqalign{
\{\rm K^0 \equiv d\bar s:0\>0\>0\>0\>0\>2\},&\>\>\{\rm \bar K^0 \equiv \bar d s:0\>0\>0\>0\>0\>\bar2\},
\>\>\{\rm K^+ \equiv u\bar s:0\>\bar 2\>0\>0\>0\>2\},\>\>\{\rm K^- \equiv s\bar u:0\>2\>0\>0\>0\>\bar2\},\cr
\>\{\rm D^+ \equiv u\bar s:0\>\bar 2\>0\>0\>\>0\bar 2\},&\>\{\rm D^- \equiv u\bar s:0\>2\>0\>0\>0\>2\},\>\>
\>\>\{\pi^- \equiv d\bar u:0\>2\>0\>0\>0\>0\},\>\>\{\pi^+ \equiv d\bar u:0\>\bar2\>0\>0\>0\>0\}.}\eqno (8.10)
$$
Quantum number conservation is demonstrated in the following two observed examples:  
$$\eqalign{
\{\rm D^+:0\>\bar 2\>0\>0\>\>0\bar 2\}\to &\{\rm K^-:0\>2\>0\>0\>0\>\bar2\}
+ \{\pi^+:0\>\bar2\>0\>0\>0\>0\} + \{\pi^+:0\>\bar2\>0\>0\>0\>0\},\cr
\{\rm D^-:0\> 2\>0\>0\>\>0 2\}\to &\{\rm K^+:0\>\bar2\>0\>0\>0\>2\} 
+ \{\pi^-:0\>2\>0\>0\>0\>0\} + \{\pi^-:0\>2\>0\>0\>0\>0\}, }	                   \eqno (8.11)
$$ 
while the following possible decays, which conserve charge, but not G, 
have not been observed: 
$$\eqalign{
	\{\rm D^+:0\>\bar 2\>0\>0\>\>0\bar 2\}\to &\{\rm K^+:0\>\bar2\>0\>0\>0\>2\} 
	+ \{\pi^+:0\>\bar2\>0\>0\>0\>0\} + \{\pi^-:0\>2\>0\>0\>0\>0\},\cr
	\{\rm D^-:0\> 2\>0\>0\>\>0 2\}\to &\{\rm K^-:0\>2\>0\>0\>0\>\bar2\} 
	+ \{\pi^+:0\>\bar2\>0\>0\>0\>0\} + \{\pi^-:0\>2\>0\>0\>0\>0\}. }	  \eqno (8.12)   
$$

$\rm K^0$ mesons are produced by the strong interaction
$$
\rm \{\pi^-:(0\>2\>0\>0\>2\>2\}+\{\rm p:3\>\bar 1\>1\>1\>\bar 1\>\bar 1\}\to 
\{\Lambda:3\>1\>1\>1\>1\>\bar 1\}+\{K^0:(0\>0\>0\>0\>0\>2\}.                               \eqno (8.13)
$$
They are of particular interest because of their involvement in
discussions of parity non-conservation (e.g. [18], \S14.4). It is relevant, 
therefore, that the above analysis shows $\rm K^0$ mesons 
to have no intrinsic parity. A detailed discussion of this problem is
given in \S4 of [20].

As fermion signatures are linearly independent is not possible to
(physically) construct a fermion in one generation from combinations of
fermions in other generations.
In particular, no interaction of fermions in generations 1-3 can produce
a 4th generation fermion. Possible ways to observe them are currently
under examination. Particles with  A=B=C=D=E=F=G=0 have no observable characteristics
other than energy, i.e. mass and momentum. They can only decay into particle/anti-particle
pairs. Important examples are the $\pi^0$ meson and the Higgs boson. 

\beginsection \S9. Results and conclusions

A description of the elementary fermions in terms of seven commuting elements
of the Clifford Unification algebra $Cl_{7,7}$ [14] has been obtained. Each of 
these defines one of the seven binary quantum numbers, A,..,G,  four of which
are defined in existing theories.  The main result is that
all seven quantum numbers are conserved in particle decays and interactions. 

Integration of the distinct applications of $Cl_{1,3}$ in the Dirac theory
and classical mechanics led, in \S2, to the identification of
$Cl_{3,3}$ as the fundamental algebra which describes the relationship between
fermion doublets and space-time and provides the foundation
of this work. Three sections establish the physical
interpretations of the quantum numbers A,B,C:

\item {\S3} describes lepton doublets, which comprise all
eight lepton states in a single generation in terms of binary quantum numbers 
A,B,C, where A specifies the spin direction for an arbitrary spatial orientation, B 
determines the coordinate time-direction and distinguishes between leptons 
and anti-leptons, and C specifies lepton intrinsic parities. 
Later sections show this analysis applies to all fermion doublets.

\item {\S4} reformulates the Dirac equation, solving the negative mass 
problem while retaining Dirac 4-spinors as free 
fermion solutions. 

\item {\S5} reformulates electro-weak interactions in a way that does
not involve chirality and shows that they necessarily
produce a parity change.

\item {\S6} The five commuting elements of $Cl_{5,5}$ 
are shown to extend the physical interpretation of its 
$Cl_{3,3}$ sub-algebra by defining quantum numbers D and E that distinguish 
quarks and leptons. Only quarks have the quantum numbers that
permit gluon interactions. 

\item {\S7} The seven commuting elements of the $Cl_{7,7}$ algebra are 
shown to extend the interpretation of $Cl_{5,5}$ to define 
the quantum numbers F and G, which distinguish four 
generations of fermions. 

The electric charges on all observed fermions 
are expressed in terms of their signature defined by the 
quantum numbers BCDEFG, showing quantum number conservation 
is in agreement with electric charge is conservation.
Fermions in the 4th generation, which has been 
identified [21] as providing the material structure of the dark matter, 
carry different electric charges to those in the 
first three generations, which compose ordinary matter. No 
fourth generation leptons have zero electric charge,
consistent with experiments showing that only three types of
neutrino exist.

\beginsection Acknowledgements

I am particularly grateful to Professor Ron King for his kind and patient help, 
given over many years, to correct my mathematical and conceptual errors. Thanks 
are also due to Professor Ian Aitchison for his patience in pointing out errors 
in some of my early assumptions.  I am also grateful to Professors Geoffrey Stedman 
and Brian Judd for their continuing encouragement and support.

\beginsection Appendix A: Properties of Clifford algebras relevant to this work

The $Cl_{p,q}$ algebra has $n=p+q$ anti-commuting generators $\gamma^i$. 
$p$ generators are square roots of unity and are referred to as {\it time-like}, 
and $q$ are square roots of $-1$ and are referred to as {\it space-like}. 
This nomenclature is also applied to all elements of the algebra constructed
from products of the generators. Eigenvalues of time-like elements
are $\pm 1$ and of space-like elements are $\pm i$.
In this work the time-like generators are labelled $i=a,b,c,$ etc.
and the space-like generators are labelled $i=1,2,3,$ etc. The unit element
is denoted ${\bf 1}$.  Products of generators are anti-symmetric in the 
exchange of any two indices, e.g.
$\gamma^i \gamma^j = {1\over 2}(\gamma^i \gamma^j - \gamma^j \gamma^i)= \gamma^{ij}
= -\gamma^{ji}$. All products are square roots of $\pm{\bf 1}$ and can be interpreted 
as unit measures of physical quantities. The product of all its generators 
closes the algebra, giving $Cl_{p,q}$ a total of $2^n = 2^{p+q}$ elements. 
Of particular interest is set of n/2 pair products of generators,
in which each generator appears once. Anti-commutation of the 
generators ensures that these products commute.

The aim of this work is to find a description of all the elementary fermions
employing an algebra in which every element has a unique physical interpretation.
The only unique element in any Clifford algebra is its 'volume' element, 
defined as the product of all $n$ generators, viz.
$$
\gamma^0 = \gamma^1\gamma^2 ....\gamma^q \gamma^a\gamma^b......\gamma^p  \eqno(A.1)
$$
where $q$ labels the $q$th space-like generator and $p$ labels 
the $p$th time-like generator. $\gamma^0$ measures the unit
of time in any coordinate system.

Clifford algebras have several general isomorphisms  (e.g. see \S16.3 
of [16]). Of particular interest are
$$
Cl_{p,q}\otimes Cl_{1,1}\simeq Cl_{p+1,q+1},     \eqno (A.2a)
$$
$$
Cl_{p,q}\simeq  Cl_{q+1,p-1},                    \eqno (A.2b)
$$
$$
Cl_{p+8,q}\simeq Cl_{p,q+8}\simeq Cl_{p,q}\otimes Cl_{8,0} \simeq Cl_{p,q}\otimes Cl_{0,8}, \eqno (A.2c)
$$
and, if $p\ge 4$,
$$ 
Cl_{p,q}\simeq Cl_{p-4,q+4}.                       \eqno (A.2d)
$$

This work is based on the properties of the $Cl_{7,7}$ 
algebra. Its paired generators
produce the seven commuting elements of the 
algebra that distinguish $2^7$ fermions and determine 
their properties. It has the isomorphisms
$$
Cl_{7,7}\simeq Cl_{11,3}\simeq Cl_{3,3}\otimes Cl_{4,4}\simeq Cl_{3,3}\otimes Cl_{2,2}\otimes Cl_{2,2}.                                                     \eqno (A.3)
$$
$Cl_{3,3}\simeq  Cl_{0,6} \simeq  Cl_{4,2}$ relates
the properties of fermion doublets to space-time. 
$Cl_{4,4}\simeq Cl_{2,2}\otimes Cl_{2,2}$ describes 
fermion properties of colour and generation,
that are unrelated to space-time.

\beginsection Appendix B: Matrix representation of $Cl_{3,3}$

The $\hat\gamma$-matrix representation of $Cl_{3,3}$
has 64 linearly independent real 8$\times$8 matrices. 
These are expressed as a 
multiplication table below, which gives the products of
the representation matrices of the elements of $Cl_{1,3}$ (left
factors) with the unit matrix and matrices of the time-like generators
$\hat\gamma^a,\>\hat\gamma^b,\>\hat\gamma^c$ (right factors).   
Each matrix is expressed as a Kronecker product of three 
real $2\times 2$ matrices defined by
$$ 
 {\bf P}= \left(\matrix{ 0 & -1\cr 1 & 0\cr} \right),\>
 \>{\bf Q}=\left(\matrix{ 0 & 1\cr 1 & 0\cr} \right),\>
 \>{\bf R}=\left(\matrix{ -1 & 0\cr 0 & 1\cr} \right),                   \eqno (B.1)                        
$$
which satisfy 
$$
-{\bf P}^2 = {\bf Q}^2 = {\bf R}^2 = {\bf I}, \>\>
{\bf P}{\bf Q} = {\bf R} = -{\bf Q}{\bf P},\>{\bf P}{\bf R} = -{\bf
	Q} = -{\bf R}{\bf P}, \>{\bf Q}{\bf R} = -{\bf P} = -{\bf R}{\bf Q}. \eqno (B.2)
$$
\vskip10pt A representation of $Cl_{3,3}$ by the $\hat{\gamma}^*$ matrices of the 
fermion coordinate frame is: 
$$\vbox
{\settabs 6 \columns
	\+|||||||||||||||||||||||||||||||||||||||\cr \+ &
	$\>\>\>\>\>{\bf 1}_3\> $  &$\>\>\>\>\>\>\>\>\>\hat\gamma^a$&
	$\>\>\>\>\>\>\>\>\>\hat\gamma^b$&$\>\>\>\>\>\>\>\>\>\hat\gamma^c $\cr
	\+|||||||||||||||||||||||||||||||||||||||\cr \+$\>\>\>\>\>\>\>{\bf
		1}_3$&$\>\>\>{\bf I}\otimes {\bf I}\otimes {\bf I}$& $\>\>\> {\bf I}\otimes {\bf Q
	}\otimes{\bf I}$& $\>\>\>i{\bf R}\otimes {\bf P}\otimes {\bf Q}$& $\>\>\>\>{\bf
		R}\otimes {\bf P}\otimes {\bf P}$\cr \+&&&\cr
	\+$\>\>\>\>\>\>\>\hat\gamma^\pi$&$\>\> i{\bf I}\otimes {\bf Q}\otimes{\bf R}$&
	$\>\> i{\bf I}\otimes {\bf I}\otimes{\bf R}$& $\>\>\> {\bf R}\otimes {\bf
		R}\otimes{\bf P}$ &$\> -i{\bf R}\otimes {\bf R}\otimes{\bf Q}$\cr
	\+|||||||||||||||||||||||||||||||||||||||\cr \+$\>\>\>\>\>\>\>\hat\gamma^{*0}
	$&$-{\bf I}\otimes {\bf R}\otimes {\bf I}$& $ -{\bf I}\otimes {\bf P
	}\otimes{\bf I}$& $-i{\bf R}\otimes {\bf Q}\otimes{\bf Q}$& $-{\bf R}\otimes
	{\bf Q}\otimes {\bf P}$\cr\+&&&\cr 
	\+$\>\>\>\>\>\>\>\hat\gamma^{*1}$&$\>\>{\bf
		Q}\otimes {\bf P}\otimes {\bf I}$& ${\bf Q}\otimes {\bf R }\otimes{\bf I}$&
	$-i{\bf P}\otimes {\bf I}\otimes {\bf Q}$& $ {\bf P}\otimes {\bf I}\otimes{\bf
		P}$\cr \+&&&\cr 
	\+$\>\>\>\>\>\>\>\hat\gamma^{*2}$&$-{\bf R}\otimes {\bf
		P}\otimes {\bf R}$& $\>\>-{\bf R}\otimes {\bf R}\otimes {\bf R}$& $\>\>\>i{\bf
		I}\otimes {\bf I}\otimes{\bf P}$&$ {\bf I}\otimes {\bf I}\otimes{\bf Q}$
	\cr\+&&&\cr 
	\+$\>\>\>\>\>\>\>\hat\gamma^{*3} $&$\>\>\> -i{\bf P}\otimes {\bf P
	}\otimes{\bf I}$& $\>\>-i {\bf P}\otimes {\bf R }\otimes{\bf I}$& $\>\>\>\>{\bf
		Q}\otimes {\bf I}\otimes {\bf Q}$&$ -i{\bf Q}\otimes {\bf I}\otimes{\bf P}$ \cr
	\+|||||||||||||||||||||||||||||||||||||||\cr
	\+$\>\>\>\>\>\>\>\hat\gamma^{*12}$&$-{\bf P}\otimes {\bf I}\otimes {\bf R}$&
	$-{\bf P}\otimes {\bf Q}\otimes{\bf R}$& $\>i {\bf Q}\otimes {\bf P}\otimes{\bf
		P}$& $\>\>{\bf Q}\otimes {\bf P}\otimes {\bf Q}$\cr\+&&&\cr
	\+$\>\>\>\>\>\>\>\hat\gamma^{*31}$&$\>\> i{\bf R}\otimes {\bf I}\otimes{\bf I}$&
	$\>\> i{\bf R}\otimes {\bf Q}\otimes{\bf I}$& $\>- {\bf I}\otimes {\bf
		P}\otimes{\bf Q}$ &$i {\bf I}\otimes {\bf P}\otimes{\bf P}$\cr\+&&&\cr
	\+$\>\>\>\>\>\>\>\hat\gamma^{*23}$&$\>\> -i{\bf Q}\otimes {\bf I}\otimes{\bf R}$&
	$ -i{\bf Q}\otimes {\bf Q}\otimes{\bf R}$& $\>-{\bf P}\otimes {\bf P}\otimes
	{\bf P}$ &$\> i{\bf P}\otimes {\bf P}\otimes{\bf Q}$\cr\+&&&\cr
	\+$\>\>\>\>\>\>\>\hat\gamma^{*03}$&$ i{\bf P} \otimes {\bf Q}\otimes{\bf I}$&
	$i{\bf P} \otimes {\bf I}\otimes{\bf I}$& $\>-{\bf Q}\otimes {\bf R}\otimes{\bf
		Q}$ &$\>i {\bf Q}\otimes {\bf R}\otimes{\bf P}$\cr\+&&&\cr
	\+$\>\>\>\>\>\>\>\hat\gamma^{*02}$&${\bf R}\otimes {\bf Q}\otimes{\bf R}$& $ {\bf
		R}\otimes {\bf I}\otimes{\bf R}$& $\>\>-i {\bf I}\otimes {\bf R}\otimes{\bf P}$
	&$\>- {\bf I}\otimes {\bf R}\otimes{\bf Q}$\cr\+&&&\cr
	\+$\>\>\>\>\>\>\>\tilde\gamma^{*01}$&$-{\bf Q}\otimes {\bf Q}\otimes {\bf I}$&
	$-{\bf Q}\otimes {\bf I}\otimes {\bf I}$& $-i {\bf P}\otimes {\bf R}\otimes{\bf
		Q}$ &$\>-{\bf P}\otimes {\bf R}\otimes {\bf P}$\cr
	\+|||||||||||||||||||||||||||||||||||||||\cr
	\+$\>\>\>\>\>\>\>\hat\gamma^{*\pi 0}$&$\>\>\>i{\bf I}\otimes {\bf P}\otimes {\bf
		R}$& $\>\>i{\bf I}\otimes {\bf R}\otimes {\bf R}$& $\>\> {\bf R}\otimes {\bf
		I}\otimes{\bf P}$ &$\>\>-i {\bf R}\otimes {\bf I}\otimes{\bf Q}$\cr\+&&&\cr
	\+$\>\>\>\>\>\>\>\hat\gamma^{*\pi 1}$&$-i{\bf Q}\otimes {\bf R}\otimes{\bf R}$&
	$\>-i{\bf Q}\otimes {\bf P}\otimes{\bf R}$& $\>- {\bf P}\otimes {\bf
		Q}\otimes{\bf P}$ &$\>-i {\bf P}\otimes {\bf Q}\otimes{\bf Q}$\cr\+&&&\cr
	\+$\>\>\>\>\>\>\>\hat\gamma^{*\pi 2}$&$\>\> i{\bf R}\otimes {\bf R}\otimes{\bf I}$&
	$\>\>\> i{\bf R}\otimes {\bf P}\otimes{\bf I}$& $\>\>\>-{\bf I}\otimes {\bf
		Q}\otimes{\bf Q}$&$\>\> i{\bf I}\otimes {\bf Q}\otimes{\bf P}$ \cr\+&&&\cr
	\+$\>\>\>\>\>\>\>\hat\gamma^{*\pi 3}$&$- {\bf P}\otimes {\bf R}\otimes{\bf R}$&
	$-{\bf P}\otimes {\bf P}\otimes {\bf R}$& $\>\>i {\bf Q}\otimes {\bf
		Q}\otimes{\bf P}$ &$\>\>{\bf Q}\otimes {\bf Q}\otimes {\bf Q}$\cr
	\+|||||||||||||||||||||||||||||||||||||||\cr} $$

Although these matrix representations are 
specifically related to fermion coordinate frames,
this table can also be used to check relationships
between $\hat\gamma$ elements of  $Cl_{3,3}$ in
any coordinate frame.   Its second and third columns
show matrix representations with block diagonal
structure, while the fourth and fifth columns have cross 
block structure. In particular, fields and Lorentz transformations have 
the block diagonal representation
$$
{\bf L}  = \hat\gamma^{\mu\nu} L_{\mu\nu} 
= \left(\matrix{ {\bf L}_a&0\cr
	0& {\bf L}_b\cr}
\right),                                                             \eqno (B.3)
$$
where
$$
{\bf 
L}_a = 
\left(\matrix{-iL_{31}&-L_{12}+iL_{23}& L_{02}        & L_{01}-iL_{03}\cr
	L_{12}+iL_{23} &iL_{31}           & L_{01}+iL_{03}&-L_{02}\cr
	L_{02}         &L_{01}-iL_{03}    & -iL_{31}      &-L_{12}+iL_{23}\cr
	L_{01}+iL_{03} &-L_{02}           & L_{12}+iL_{23}& iL_{31} \cr}\right)  \eqno (B.3a)        
$$
and
$$
{\bf L}_b =
\left(\matrix{-iL_{31}&L_{12}-iL_{23}& -L_{02}       & L_{01}-iL_{03}\cr
	-L_{12}-iL_{23}   &iL_{31}       & L_{01}+iL_{03}&L_{02}\cr
	-L_{02}           &L_{01}-iL_{03}& -iL_{31}      &L_{12}-iL_{23}\cr
	L_{01}+iL_{03}    &L_{02}        &-L_{12}-iL_{23}& iL_{31} \cr}\right).  \eqno (B.3b)      
$$

Real diagonal matrices that correspond 
to the quantum numbers A,B,C are
$$
\hat\gamma^{\rm A}= i\hat\gamma^{12}\equiv diag\{1 \bar1 1 \bar1 1 \bar1 1 \bar1\},\>\> 
\hat\gamma^{\rm B}= \hat\gamma^0 \equiv diag\{1 1 \bar1 \bar1 1 1 \bar1 \bar1\},\>\>
\hat\gamma^{\rm C} =i\hat\gamma^{bc}\equiv diag\{1 1 1 1 \bar1 \bar1 \bar1 \bar1\}.       \eqno (B.4)
$$ 
These define the projection operators that distinguish the eight
states of fermion doublets, viz.
$$
P(lepton)={1\over 8}({\bf 1}\pm \hat\gamma^{\rm A})({\bf 1}
\pm \hat\gamma^{\rm B})({\bf 1} \pm \hat\gamma^{\rm C}).                    \eqno (B.5)
$$

\vfill\eject

\beginsection References

\frenchspacing

\item {[1]} Benn, I.M. and Tucker, R.W. (1988) An Introduction to Spinors and Geometry with 
Applications in Physics (Adam Hilger)

\item {[2]} Doran, Chris and Lasenby, Anthony  (2003) Geometric Algebra for Physicists
(Cambridge University Press) 

\item {[3]} Eddington, Sir A.S. (1946) Fundamental Theory (Cambridge University Press)

\item {[4]} Trayling, Greg. (1999) A Geometric Approach to the Standard Model {arXiv: hep-th:9912231v1} 

\item {[5]} Trayling, Greg and Baylis, W.E. (2001) A geometric basis for the standard-model
gauge group. {\it J. Phys. A: Math. Gen.} {\bf 34}, 3309-3324 {arXiv: hep-th:0103137v2}

\item{[6]} Furey, C. (2018) Three generations, two unbroken gauge symmetries, and one 
eight-dimensional algebra Phys. Lett. B {\bf 785} 84 {arXiv:1910.08395v?}

\item {[7]} Dartora, C. A. and Cabrera, G. G. (2009) The Dirac equation and a non-chiral electroweak theory
in six dimensional space-time from a locally gauged $SO(3,3)$ symmetry group {arXiv: 0901.4230v1}
Int J Theor Phys (2010) {\bf 49}:51-61 

\item {[8]} Stoica, O. C. (2018) The Standard Model Algebra: 
Leptons, Quarks and Gauge from the Complex 
Clifford Algebra Cl$_6$ {\it Adv. Appl. Clifford Algebra}
{\bf 28}, 52. {arXiv:1702.04336v3} 

\item {[9]} Yamatsu, Naoki (2020) USp(32) Special Grand Unification {arXiv:2007.08067v1}  

\item{[10]} Gording, Brage and Schmidt-May, Agnis. (2020) The Unified Standard Model {arXiv:1909.05641}

\item {[11]} Pav\v si\v c, Matej (2021) Clifford Algebras, Spinors and Cl(8,8) Unification {arXiv:2105.11808}

\item {[12]} Patel, Aditya Ankur and Singh, Tejinder P (2023) CKM matrix parameters 
from an algebra\vskip 1pt {arXiv:2305.00668}

\item {[13]} Bhatt, Vivan; Mondal, Rajrupa; Vaibhav, Vatsalya and Singh, Tejinder P. (2023) \vskip 1pt
Majorana Neutrinos, Exceptional Jordan Algebra and Mass Ratios for Charged Fermions \vskip 1pt 
J.Phys.G::Nucl. Part. Phys. {\bf 49} 045007 (2022) {arXiv:2108.05787} 

\item {[14]} Newman, Douglas (2023) Unified theory of elementary fermions and their interactions
based on Clifford algebra {arXiv: 2108.08274v11}

\item {[15]} Newman, Douglas (2023) Problems with the Standard Model of particle physics {arXiv: 2308.12295v2}

\item {[16]} Lounesto, Pertti (1997) Clifford Algebras and Spinors (Cambridge University Press) 

\item {[17]} Vax,Jr., Jayme and da Rocha,Jr., Rold$\rm \tilde a$o (2016) An Introduction to Clifford
Algebras and Spinors (Oxford University Press)

\item {[18]} Thomson, Mark. (2013) Modern Particle Physics (Cambridge University Press)

\item {[19]} Bettini, Alessandro (2008) Introduction to Elementary
Particle Physics (Cambridge University Press) 

\item {[20]} Newman, Douglas (2024) Quantum number conservation in meson interactions {arXiv: 2403.07962}

\item {[21]} Newman, Douglas (2022) Fourth generation fermions providing new candidates
 for dark matter and dark energy {arXiv: 2201.13238}
\vskip5pt  

 \end
 \bye